\documentclass[a4paper,11pt]{article}
\pdfoutput=1
\usepackage[utf8]{inputenc}
\usepackage[T1]{fontenc}

\usepackage{textgreek}
\usepackage[svgnames]{xcolor}
\usepackage{jinstpub}

\usepackage{wasysym}
\usepackage{xspace}				% taking care of spaces after command.
\usepackage[nameinlink, noabbrev, capitalise]{cleveref}           % use cref instead of autoref
\usepackage{tikz}
\tikzset{
	hint/.style = {
		black,shorten <=2pt, postaction={
			draw,latex-,line width=2pt,white,shorten <=0pt,postaction={
				draw,latex-,black,line width=1pt,shorten <=2pt,
			},
		},
	},
}

\tikzset{
	hint2/.style = {
		black,shorten <=2pt, shorten >=2pt, postaction={
			draw,latex-latex,line width=2pt,white,shorten <=0pt,shorten >=0pt,postaction={
				draw,latex-latex,black,line width=1pt,shorten <=2pt, shorten >=2pt,
			},
		},
	},
}

\tikzset{
	hintnb/.style = {latex-,black,line width=1pt}
}
\tikzset{
	hint2nb/.style = {latex-latex,black,line width=1pt}
}

\crefformat{equation}{#2(#1)#3}

\usepackage{siunitx}
\usepackage{layouts}

\newcommand{\Kr}[1][83m]{$^\mathrm{#1}$Kr\xspace}
\newcommand{\dV}{\ensuremath{\Delta V_\text{ag}}\xspace}

\newcommand{\mus}{\micro \second}
\newcommand{\mum}{\micro \meter}
\newcommand{\pepph}{PE \per photon}
\newcommand{\pepel}{PE \per electron}

\newcommand{\nel}{N_\mathrm{e}}
\newcommand{\nph}{N_\mathrm{\upgamma}}

\title{Proportional scintillation in liquid xenon: demonstration in a single-phase liquid-only time projection chamber}

\author{Florian~Tönnies,}
\author{Adam~Brown,}
\author{Baris~Kiyim,}
\author{Fabian~Kuger,}
\author{Sebastian~Lindemann,}
\author{Patrick~Meinhardt,}
\author{Marc~Schumann,}
\author{Andrew~Stevens}

\affiliation{Physikalisches Institut, Universität Freiburg, 79104 Freiburg, Germany}

\emailAdd{florian.toennies@physik.uni-freiburg.de}
\emailAdd{adam.brown@physik.uni-freiburg.de}
\emailAdd{marc.schumann@physik.uni-freiburg.de}

\abstract{
The largest direct dark matter search experiments to date employ dual-phase time projection chambers (TPCs) with liquid noble gas targets. These detect both the primary photons generated by particle interactions in the liquid target, as well as proportional secondary scintillation light created by the ionization electrons in a strong electric field in the gas phase between the liquid-gas interface and the anode.
In this work, we describe the detection of charge signals in a small-scale single-phase liquid-xenon-only TPC, that features the well-established TPC geometry with light readout above and below a cylindrical target. In the single-phase TPC, the proportional scintillation light (S2) is generated in liquid xenon in close proximity to \qty{10}{\mum} diameter anode wires. 
The detector was characterized and the proportional scintillation process was studied using the \qtylist{32.1;9.4}{keV} signals from \Kr decays.
A charge gain factor~$g_2$ of up to \qty{1.9(3)}{\pepel} was reached at an anode voltage \qty{4.4}{kV} higher than the gate electrode \qty{5}{mm} below it, corresponding to \qty{29(6)} photons emitted per ionization electron.
The duration of S2 signals is dominated by electron diffusion and approaches the xenon de-excitation timescale for very short electron drift times. 
The electron drift velocity and the longitudinal diffusion constant were measured at a drift field of \qty{470}{V/cm}. The results agree with the literature and demonstrate that a single-phase TPC can be operated successfully.
}

\keywords{Time projection chamber, liquid noble gas detectors, liquid xenon, electroluminescence}

\begin{document}
\maketitle
\flushbottom

\section{Introduction}
\label{sec:intro}

Dual-phase (liquid/gas) time projection chambers (TPCs) filled with the liquefied noble gases xenon or argon are widely used in low-background experiments searching for low-energy rare events such as WIMP dark matter~\cite{appec_report}. Currently operating detectors with active targets above the tonne scale are PandaX-4T \cite{result_pandax4t_meng}, XENONnT \cite{experiment_XENONnT_wimp_sensitivity} and LZ \cite{experiment_lz}, all employing liquid xenon (LXe) targets.

In dual-phase xenon TPCs~\cite{Schumann:2014uva}, particle interactions in the LXe target excite and ionize xenon atoms. Subsequent de-excitations lead to a prompt scintillation light signal (S1), that is detected by light sensors installed above and below the cylindrical TPC. Ionization electrons are drifted across the LXe target in an electric field of typically around \qty{100}{V/cm}. This is established between a negatively biased cathode, installed below the LXe target, and a gate electrode (typically at ground potential), just below the liquid-gas interface. A second, stronger electric extraction field ($E\sim\qty{10}{kV/cm}$) is established between the gate electrode and the positively biased anode, a few millimeters above the liquid surface. This field extracts the electrons into the gas phase where they create a secondary light signal (S2), proportional in size to the number of electrons. 
The photons from both processes release photoelectrons (PE) in the light sensors that create the recorded signal.
The ratio of detected photoelectrons to photons produced in the S1 is given by the gain factor $g_1$.
The electroluminescence yield describes the average number of secondary photons created by a single electron extracted into the gas. The gain factor $g_2$ combines this with the photon detection efficiency to give the number of detected photoelectrons per electron.
Typical $g_2$ values of current dual-phase experiments range from about 17~to almost \qty{60}{PE/electron} \cite{result_xenonnt_lowER,XENON:2021qze,LZ:2023poo}. The position of the primary interaction in the TPC can be inferred from the pattern of the detected S2~signal across the top photosensors ($xy$) and by the time difference between S1 and S2~signals
($z$). The number of individual S2~signals, indicating the scatter multiplicity, and the ratio~S1/S2 can be used to separate dark matter signals from background events~\cite{Aprile:2006kx}.

Although dual-phase LXe TPCs currently provide the tightest constraints on WIMP dark matter interactions for WIMP masses above about \qty{3}{GeV/c^2}~\cite{appec_report}, the technology faces experimental challenges.
All TPC electrodes have to be highly optically transparent to enable light detection; they are typically made from individual parallel wires~\cite{experiment_xenon1t,XENON:2024wpa}, two-dimensional (etched or woven) meshes~\cite{experiment_xenon100,experiment_lz}, or -- in case of cathode and anode -- solid quartz-plates with a conductive layer~\cite{DarkSide:2014llq}. The size of the S2~signal depends on the electron path length in the xenon gas and the local electric field, and no S2~signal is produced in regions where the anode touches the liquid-gas interface.
This means the gate electrode and anode have to be precisely positioned, parallel to each other and to the liquid-gas interface, for a uniform detector response. The liquid level also needs to be kept stable over long time periods.
Due to the combination of electrostatic and gravitational forces, a position-dependent deflection of the anode and gate planes cannot be avoided, which requires position-dependent corrections to the S2~signal~\cite{experiment_xenon100,experiment_xenon1t}. To minimize this effect and prevent the anode from touching the liquid-gas-interface, the electrodes wires or meshes must be tensioned. This implies more massive support frames for the wires, leading to increased radioactivity and associated background.

Creating the proportional S2~scintillation signal in the liquid xenon phase overcomes these issues.
However, an electric field greater than $\sim$\qty{400}{kV/cm} is required for the electron to excite xenon atoms in the liquid phase. Such fields can be created in the $E(r) \propto 1/r$ radial field close to the surface of thin anode wires~\cite{Ye:2014gga}. However, excessively high fields result in electron multiplication, worsening the energy resolution~\cite{singlephase_tpc_kuger}.

Proportional scintillation in the liquid phase also removes the need to extract electrons into the gaseous phase. This means that there is no delayed extraction of electrons, which contribute significantly to the accidental coincidence background of current dual-phase TPCs~\cite{XENON:2021qze,result_xenonnt_firstdmsearch}.
Due to the absence of the liquid-gas interface, total internal reflection on the liquid-gas interface is avoided, increasing the S1 photon detection efficiency by about~10\%~\cite{simulation_chroma_paper}.

Proportional scintillation in liquid xenon was first observed in proportional scintillation counters around thin wires~\cite{Lansiart1976,Masuda:1978tjp}. More recently, the technology regained attention for dark matter searches~\cite{Ye:2014gga}. Aprile et al.~studied \qty{5.4}{MeV} signals from $^{210}$Po$~\alpha$-decays in a small cubic TPC prototype with a 5\,mm drift region and a single anode wire; the light was recorded by two photomuliplier tubes, one above and one below the target~\cite{singlephase_tpc_elena}.
Charge yields of up to \qty{1.8}{\pepel} have been achieved in a novel geometry: the radial TPC~\cite{radial_tpc_lin}, where a single thin anode wire is centrally located in a cylindrical liquid xenon volume~\cite{radial_tpc_lin,radial_tpc_qi}.

In this work we demonstrate proportional scintillation in liquid xenon in a standard cylindrical TPC, i.e., a small-scale version of current dark matter TPCs, which has previously been operated and characterized in dual-phase mode~\cite{experiment_xebra}. The etched hexagonal anode mesh of the dual-phase detector was replaced by a set of parallel, \qty{10}{\um} diameter wires, while the rest of the TPC remained untouched.

In \cref{sec:TPC} of this work we describe the design and operation of the small-scale single-phase TPC on the XeBRA detector platform. This provides a cryostat and a cooling system that can accommodate LXe detectors of a few kilograms, a system for gas storage and purification, as well as data acquisition (DAQ) and slow control systems. The detector is characterized and the proportional scintillation in liquid xenon is studied using low-energy \Kr-events. The data analysis and signal corrections are presented in \cref{sec:dataanalysis}.
We report on our results, in particular the electroluminescence gain and the duration of the observed S2 signals, in \cref{results}.

\raggedbottom
\section{Design and operation of the single-phase TPC}
\label{sec:TPC}

\subsection{Time Projection Chamber}

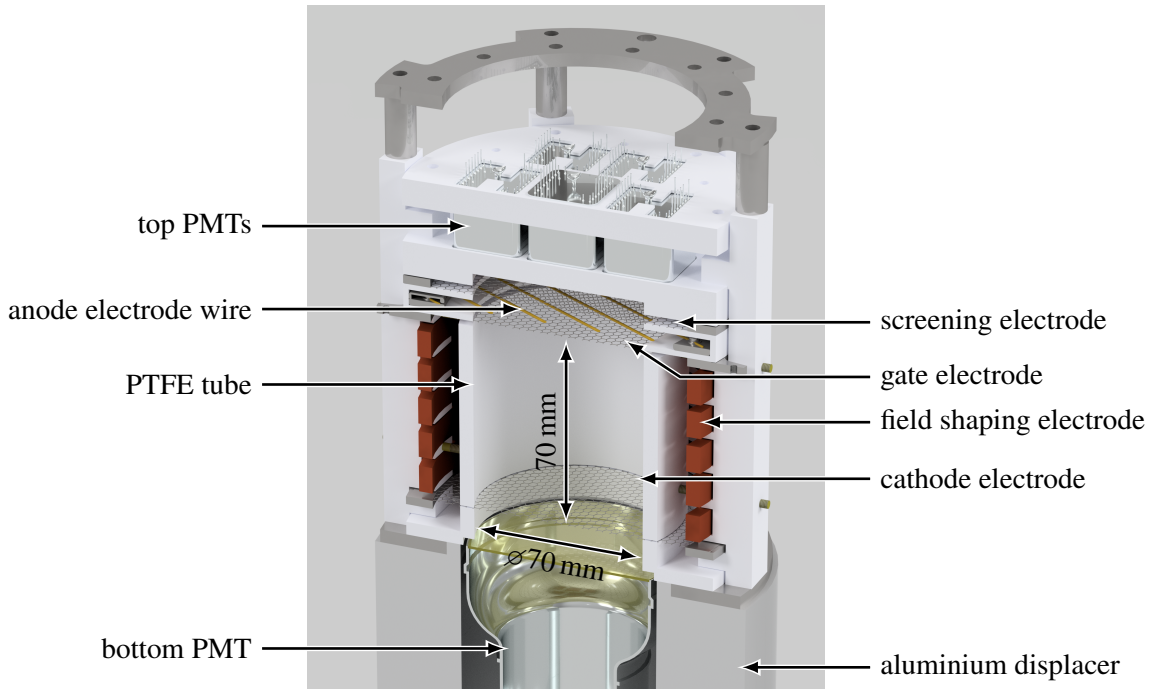
\begin{figure}
	\centering
	\begin{tikzpicture}
		\draw (0, 0) node[inner sep=0] {\includegraphics[width = .45\linewidth]{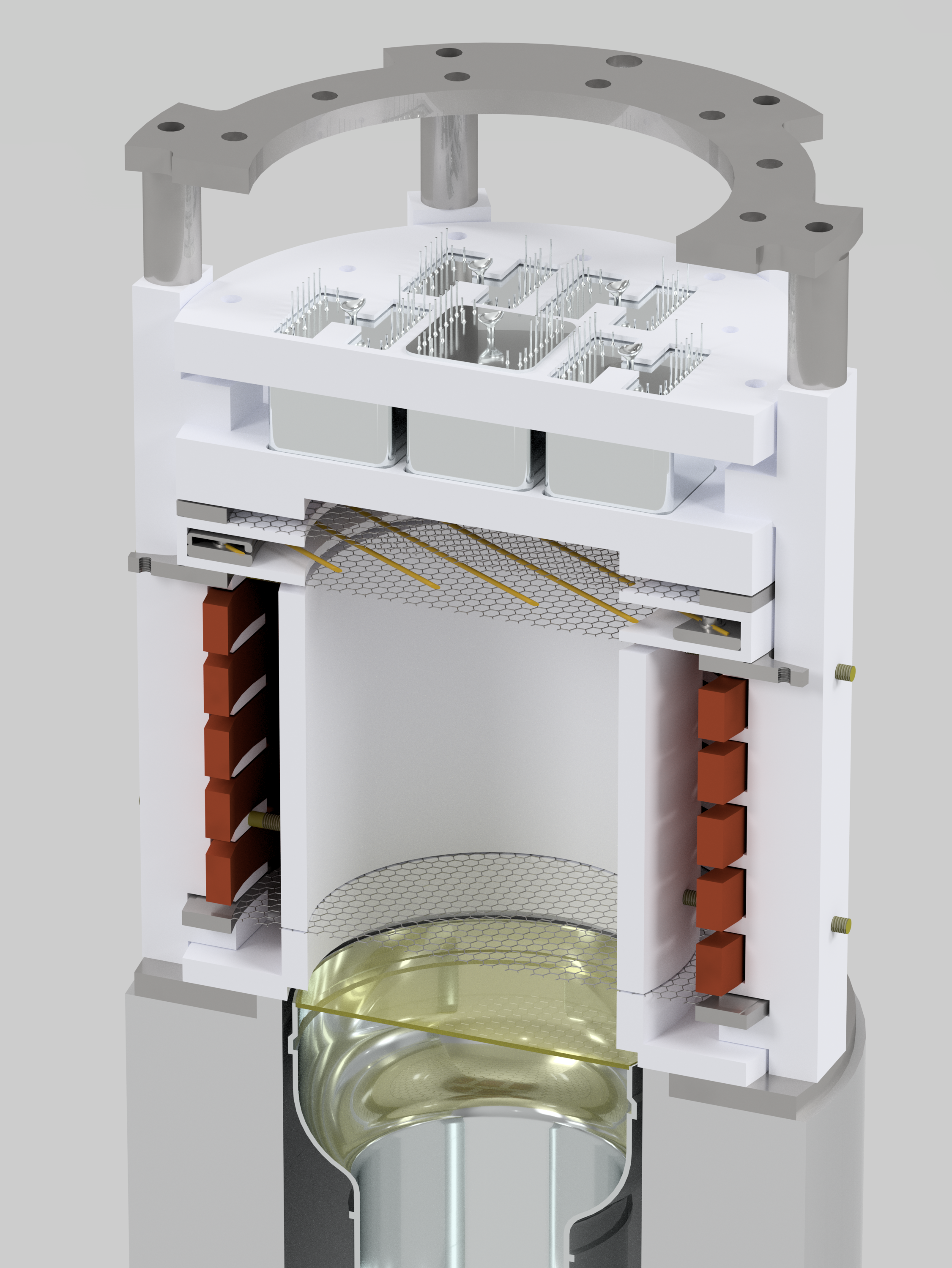}};
				
		% tubes
		\draw [hint] (-1.25, 1.6) -- (-4, 1.6) node[anchor = east, align = right] {top PMTs};
		\draw [hint] (-1.2, -.5) -- (-4, -0.5) node[anchor = east, align = right] {PTFE tube};
		\draw [hint] (-.75, -4) -- (-4, -4) node[anchor = east, align = right] {bottom PMT};
		
		% other
		\draw [hint] (2.25, -4.25) -- (4, -4.25) node[anchor = west, align = left] {aluminium displacer};
		
		% electrodes and field shaping rings
		\draw [hint] (1.4, 0.325) --  (4, 0.325) node[anchor = west, align = left] {screening electrode};
		\draw [hint] (0.75, 0.1) -- (1.50, -0.4) -- (4, -0.4) node[anchor = west, align = left] {gate electrode};
		\draw [hint] (-0.55, 0.5) -- (-4, 0.5) node[anchor = east, align = left] {anode electrode wire};

		\draw [hint] (1.75, -1) -- (4, -1) node[anchor = west, align = left] {field shaping electrode};
		\draw [hint] (0.9, -1.75) -- (4, -1.75) node[anchor = west, align = left] {cathode electrode};
		
		% distances
		
		\draw[hint2] (-1.2, -2.4) -- (1.0, -2.8) node[midway,below,sloped]{\diameter \qty{70}{mm}};
		\draw[hint2] (0, -2.35) -- (0, 0.1) node[midway,above,sloped]{\qty{70}{mm}};
		
	\end{tikzpicture}
	\caption{
		3D-rendering of the single-phase TPC. The dimensions are specified for room temperature. The anode wires of \qty{10}{\um} diameter are enlarged for visibility.
	}
	\label{fig:tpc_cad}
\end{figure}

The dual-phase TPC presented in \cite{experiment_xebra} was modified to operate it in single-phase mode, and is shown in \cref{fig:tpc_cad}. The TPC is contained in a vacuum-insulated double-walled cryostat, filled with about 10\,kg of xenon.
Roughly 0.75\,kg of LXe are contained in the cylindrical active TPC volume, where interactions can be detected.
This volume has a height of \qty{70}{mm}, between the cathode and gate electrode, and an inner diameter of \qty{70}{mm}. These dimensions reduce to \qty{69}{mm} at LXe temperature.
One Hamamatsu R11410-10 PMT~\cite{Baudis:2013xva} of 3-inch diameter is installed below the active volume. The PMT is surrounded by a solid aluminium displacer to reduce the total required amount of LXe. Seven 1\,$\times$\,1-inch Hamamatsu R8520 PMTs detect the light distributed across the top of the TPC which can be used to infer the horizontal $xy$-position of an interaction from the S2-signal, as described in \cref{sec:posrec}.

\begin{figure}
\centering
\includegraphics[width=.80\linewidth]{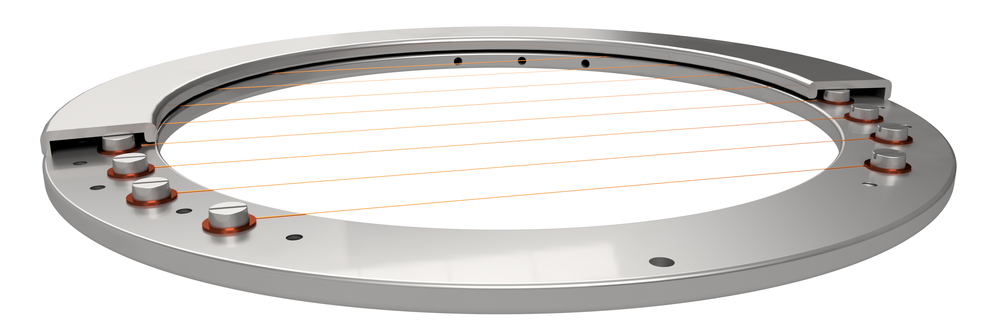}
\caption{
CAD image of the anode electrode. The gold-plated tungsten wires of \qty{10}{\mum} diameter are fixed to the stainless steel support frame with a pitch of \qty{10}{mm} using copper washers and stainless steel bolts. To increase visibility, the wire diameter is enlarged in the image.
Only half of the stainless steel piece to cover the screws is shown.
}
\label{fig:wires_cad}
\end{figure}

The anode electrode for the single-phase operation consists of a set of parallel California Fine Wire gold-plated tungsten wires of \qty{10}{\mum} diameter. This diameter leads to surface electric fields of \qtyrange{730}{1220}{kV\per cm} at anode voltages of \qtyrange{3}{5}{kV} relative to the surroundings.
The wires were stretched across a circular stainless steel frame by fixing one end of the wire between the frame and a copper washer using stainless steel M2.5 bolts, see \cref{fig:wires_cad}. The other end was tensioned with a \qty{20}{g} weight and then also fixed.
The resonant modes of all wires on the finished anode were measured and from these the individual wire tensions were found to  range from \qtyrange{30}{180}{mN}. No clear difference between the performance of the differently-tensioned individual wires was observed.

While the anode frame can accommodate wires at a pitch of 5\,mm, only every second wire was installed, since this is expected to lead to higher and more regular fields around the anode wires~\cite{singlephase_tpc_kuger}. The anode wire pitch is thus 10\,mm and six anode wires cover the cross section of the TPC.
A stainless steel cover placed on top of the bolts avoids high-field regions around sharp edges.

The gate and screening electrodes are installed 5\,mm below and above the anode, respectively, with precision-machined PTFE rings as spacers. The cathode, installed 70\,mm below the gate electrode, and the gate and screening electrodes are hexagonal stainless steel meshes.
These are etched from a \qty{150}{\mum} thick stainless steel sheet and feature a \qty{3}{mm} pitch and a strand width of \qty{150}{\mum}. The voltages of cathode, gate electrode and anode can be set independently to establish the electric drift and proportional scintillation fields.
The active volume of the TPC is enclosed by a PTFE tube of 70\,mm inner diameter.
A set of five copper field shaping electrodes, connected by high-ohmic resistors, are located \qty{10}{mm} outside the inner surface of the PTFE tube.
This separation, relatively large compared to the radius of the TPC, improves the field homogeneity inside the TPC.
Further details about the detector and the XeBRA platform can be found in~\cite{experiment_xebra}.

\newpage
\subsection{Operation}

The cryostat was filled with about 10\,kg of liquid xenon such that the top PMTs were partially submerged. For single-phase operation, neither precise liquid level control nor leveling are required. The detector was operated stably using a liquid-nitrogen based cooling system~\cite{experiment_xebra} with the gaseous xenon being kept at \qty{1.7}{bar} absolute pressure and a temperature of \qty{174}{K}.
During operation the xenon was purified by a SAES MonoTorr PS3-MT3-R-2 hot getter. Liquid xenon was extracted from the region outside the TPC via a custom-built heat exchanger and pushed back into the TPC below the cathode. Any returning gaseous xenon which was not liquefied in the heat exchanger was directed to the cold finger located in the gas phase above the detector to be liquefied. The purification system also allowed injecting \Kr atoms into the TPC for detector calibration.

The PMT signals were amplified by a factor of~10 by custom amplifiers built for the XENONnT experiment. The data acquisition system is also based on the triggerless system developed for XENON~\cite{XENON:2019bth,XENON:2022vye}. It independently digitizes every PMT waveform exceeding a threshold of \qty{28}{mV} (equivalent to around \qty{1}{PE}) using a CAEN V1724 ADC with 14\,bit resolution over a \qty{2.25}{V} dynamic range and \qty{100}{MHz} sampling frequency. 
Every PMT waveform includes 50~samples from before the threshold is crossed and 40~samples after dropping below the threshold again. One of the 7~PMTs in the top array did not operate at cryogenic temperatures, most likely due to a faulty cable connection.
The Doberman slow control system~\cite{software_doberman_2016}, specially developed for such small- to medium-scale experiments, was used to operate the system stably for several weeks.

Data were taken with varying amplification fields for proportional scintillation.
The gate and screening electrodes were always held at the same voltage as each other.
The voltage difference~\dV between the anode and the gate and screening electrodes was increased in steps of \qty{200}{V} from \qty{3.0}{kV} to \qty{5.0}{kV}.
For each value of \dV, data were recorded with two different gate and screening electrode voltages of \qty{-1}{kV} and \qty{-2}{kV}.
As no difference could be observed in the data from each of these two absolute voltage settings, all data  for a given~\dV are combined in the analysis presented here.
The cathode voltage was set to either \qty{-4.5}{kV} or \qty{-5.5}{kV} to maintain a potential difference of 3500\,V across the 69\,mm long drift region. The electric drift field~$E_d$ inside the active region was calculated using finite element simulations, taking into account the detailed geometry. The median field strength is $E_d=\qty{473}{V/cm}$, with a one-sigma spread of \qty{8}{V/cm}.
The drift time of electrons produced at the cathode was \qty{43.2(0.6)}{\mus}. This is shorter than the average electron lifetime $\tau_e \approx\qty{80}{\mus}$ caused by the residual electronegative impurities in the LXe (see~\cref{sec:dataanalysis}), meaning most electrons released in an interaction reach the anode.

Data could be acquired at eight different amplification fields~\dV. Below $\dV=\qty{3.4}{kV}$, very few S2s could be identified due to their small signal, leading to a large statistical uncertainty.
Above $\dV=\qty{4.4}{kV}$ a constant, high rate of small signals of unknown origin occupied the digitizer channels and prevented the recording of a sufficiently large number of events with identifiable S1-S2 pairs.
A previous attempt to operate the single-phase TPC was limited to even lower anode voltages. Before taking the data used in this work, a careful cleaning campaign was followed in an ISO-6 cleanroom to remove dust from the TPC.
Nevertheless, since the XeBRA facililty itself is not in a cleanroom, some dust exposure was unavoidable and this might be the cause of the high rate of small signals.
\section{Data analysis}
\label{sec:dataanalysis}

\begin{figure}
\centering
\includegraphics{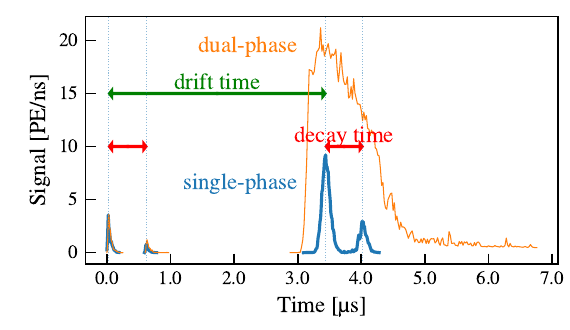}
\caption{
Example \Kr events measured with the single-phase TPC (blue) and the same TPC operated in dual-phase mode (orange, \cite{experiment_xebra}). For direct comparison events with similar drift and decay times were selected and plotted together. The decay time is indicated by the red arrows, the drift time is marked by the green arrow. 
Both signals feature the characteristic double-S1 pattern of the \Kr decays while the corresponding two S2 signals can only be clearly identified in the single-phase TPC. This is due to the different mechanisms to produce the proportional scintillation S2~light.
}
\label{fig:waveform_overview}
\end{figure}

The TPC was characterized and the production of proportional S2 signals in the liquid xenon phase was studied using a \Kr conversion electron source~\cite{Manalaysay:2009yq,Kastens:2009rt}. 
It decays in two steps, which produce electronic recoil signals of \qty{32.1}{keV} and \qty{9.4}{keV} energy and have mean lifetimes of $\tau_1=\qty{2.64(2)}{h}$ and $\tau_2=\qty{226(7)}{ns}$, respectively. The delayed-coincidence decay provides a signature which is easily identified in an unshielded, high background environment. Both decays produce an S1 with an accompanying S2, so that four signals are visible in the TPC. The time difference between the S1 and the S2 from each decay depends on the depth of the decay in the TPC. The time difference between the two S1~peaks (and the two S2~peaks) is given by the decay time of the intermediate \Kr state.

Two example \Kr events with this structure, acquired during the dual- and the single-phase operation of the same TPC, are shown in \cref{fig:waveform_overview}.
The S2 signals from the single-phase TPC are much shorter, because the S2 light is created only close to the anode wires~\cite{singlephase_tpc_kuger}. In this example, the two dual-phase S2~signals are merged due to their duration. More details on S2 duration in the single-phase TPC are presented in \cref{ssec:S2_width}.

\subsection{Event selection}
The strax framework~\cite{straxrepo}, developed by the XENON dark matter collaboration~\cite{result_xenonnt_firstdmsearch}, is used to process the raw data.
In this framework, a hit is an excursion of the digitized waveform from a single PMT above a pre-defined threshold, which is set at 28 mV
(equivalent to around 1 PE) above the baseline. The area of a hit is determined by integrating its waveform, after subtracting the baseline. This area is expressed in photoelectrons (PE) by dividing by the gain of the PMT. The statistical method described in~\cite{Saldanha:2016mkn} was used to monitor the gain in regular calibrations, in which the emission of a few PEs was stimulated by pulsed blue LED light.
Hits from any of the eight PMTs are merged into peaks observed in the TPC whenever they are within \qty{150}{ns}.
Peaks occurring within \qty{120}{\mus} of each other are grouped into an event.
This interval is chosen such that all peaks resulting from a single interaction are in the same event.
At this stage, the time of a peak is taken to be its median, or the time by which the half the total area has been detected.

\begin{figure}[t]
\centering
\includegraphics{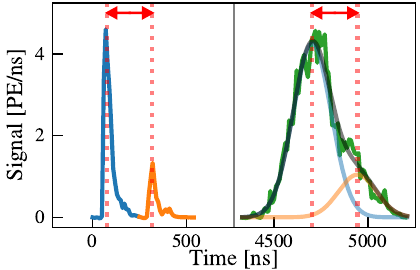}
\hfill
\includegraphics{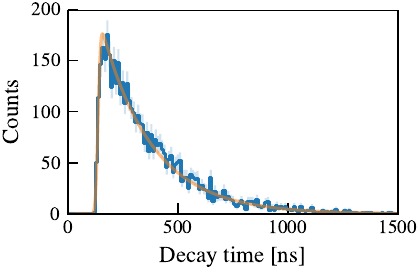}
\caption{
\textbf{Left:} S2 fitting procedure applied to a \Kr decay where the two S2 signals have significant overlap. The time difference between the two Gaussians used to fit the combined S2 peak is fixed to the time difference between the two S1s.
\textbf{Right:} The decay time distribution of the \Kr  events selected for the analysis presented in this work. A fit of an exponential function multiplied by an error function representing the peak-separating efficiency to the data yields a \Kr lifetime of \qty{230(5)}{ns}. The mid-point of the error function is at \qty{139.6(12)}{ns} and corresponds to the minimum decay time to separate the two S1 peaks.
}
\label{fig:double_gaussian_fit}
\label{fig:kr_decaytime}
\end{figure}

For the analysis presented here, signal purity is more important than statistics.
Only events with at least three peaks with areas above \qty{25}{PE} are considered as potential \Kr events and used for further analysis. The first two peaks in an event are always attributed to the two \Kr S1s, the following peaks to S2s. 
If only three peaks are present in an event, it is assumed that the two~S2s were merged into a single peak due to a combination of a short decay time and electron diffusion.
If four or more peaks are present, the third and fourth peak are assumed to be the two S2s, provided that the time between them is not more than \qty{50}{ns} different from the time between the two S1s.
If the time differs by more than this, the fourth peak is assumed to be a spurious signal such as an afterpulse, and the third peak is considered to be the two combined S2s.
Any further peaks after these are ignored.
In all cases, a sum of two Gaussian functions is fitted to the waveform containing the S2 candidates. An example of such a fit is shown in \cref{fig:double_gaussian_fit}, left. The time between the centers of the Gaussians is fixed to the decay time obtained from the S1 peaks.
The areas and widths of the Gaussians are left free in the fit.
The fit is used to obtain the individual S2 areas, their durations (signal widths), and the event's drift time, given by the temporal distance of the first S1 peak to the mean of first S2 peak.
The S2 peaks' times are redefined at this stage to be the mean of the fitted Gaussian, while the S1 peaks' times are left unchanged.
This fitting procedure enables the use of events where the S2s are relatively close in time. In these cases a more conventional approach, where the peak would be split into two parts at a certain time, could result in biased estimates of the individual S2s' areas.

Several criteria ensure the purity of the \Kr sample. Both S1s are required to have an area between~\qtylist{25;250}{PE}.
The S2 area must be between \qtylist{25;60000}{PE}, which is loose enough to avoid rejecting good events at high or low anode voltages \dV.
No further selection criteria are applied on the S2 signals.
Additionally, the time difference between the two krypton decays must be smaller than \qty{1.5}{\mus}, which cuts only 0.1\% of all krypton decays.
No lower threshold is set for the time difference. However, due to the duration of the individual S1~signals, only \Kr events with a decay time of at least \qty{140}{ns} can be separated by our analysis procedure, as seen in \cref{fig:kr_decaytime}, right.
This means that only around half of all \Kr events are used for the analysis. \cref{fig:kr_decaytime}, right, also demonstrates the purity of the selected \Kr sample: the lifetime extracted from the exponential fit is \qty{230(5)}{ns}, which agrees with the literature value of \qty{226.2(7)}{ns}~\cite{lifetime_kr_source}.

\subsection{Fiducialization}
\begin{figure}
\centering
\includegraphics{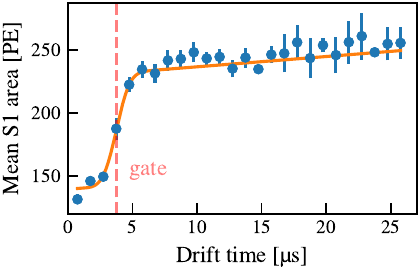}
\hfill
\includegraphics{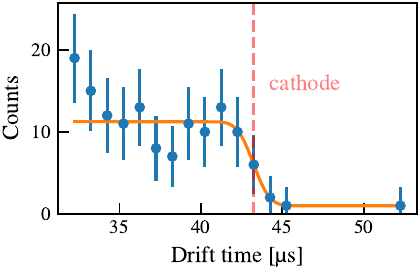}
\caption{
The drift time of events from the gate electrode (left) and cathode (right) are obtained by identifying the steps in the S1 signal size and count rate, respectively. 
In each case, an error function is fitted to the binned data and the midpoint is taken to correspond to the drift time from that electrode. In the case of the gate electrode this is $T_\text{g}=\qty{3.8(1)}{\mus}$, and for the cathode it is $T_\text{c}=\qty{43.2(6)}{\mus}$.
}
\label{fig:anode_gate_search}
\end{figure}

A homogeneous drift field is required to ensure uniform signal generation. Leakage through the cathode and gate electrodes affects the field in their vicinity. To cut these regions from the data, the procedure presented in~\cite{experiment_xebra} is used to identify the drift times corresponding to the gate electrode and cathode positions. 
As the electric field is different below and above the gate electrode, recombination and therefore the S1 area are different. This effect is used to identify the drift time of the gate electrode, as seen in \cref{fig:anode_gate_search} (left), where the mean area of the S1 signal is shown as a function of the drift time. The dependence is described by an error function multiplied by a linear term to represent the depth-dependent light collection efficiency. The drift time at 50\% of the error function is taken as the drift time of the gate electrode~$T_\text{g}$. 
Below the cathode, the field is reversed and therefore no S2 signals are seen from decays occurring there.
The drift time of the cathode~$T_\text{c}$ is found by fitting the number of events as a function of the drift time with an error function, as seen in \cref{fig:anode_gate_search} (right).

The drift time is corrected by subtracting $T_g$, to obtain only the drift time within the region below the gate electrode. Effects which occur while electrons drift from the gate electrode to the anode are considered to be an intrinsic part of the S2 light generation process.
Fiducialization in depth ($z$) is performed by keeping only events with corrected drift times between \qtylist{5;35}{\mus}, corresponding to \qtylist{8.8;61.6}{mm} below the gate electrode. A radial fiducial cut is not applied for this analysis. This is because the drift field is uniform up to the TPC radius of \qty{35}{mm} due to the large \qty{10}{mm} distance between the inner PTFE surface and the field shaping electrodes~\cite{experiment_xebra}.

\newpage
\subsection{Corrections}

\begin{figure}
\centering
\includegraphics{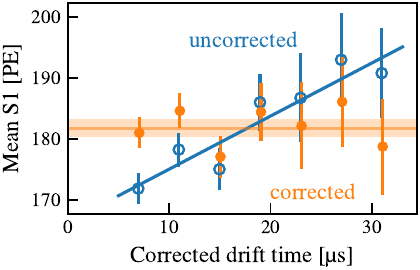}
\hfill
\includegraphics{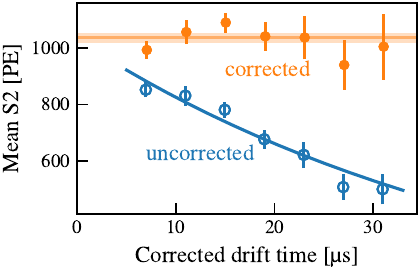}
\caption{%
Corrections of the S1 and S2 signals. 
\textbf{Left:} Monoenergetic S1 signals from the \qty{32.1}{keV} \Kr decay are impacted by the $z$-dependent light collection efficiency, which can be described by a linear function (blue). The function is used to correct the S1 areas towards the center of the TPC volume (orange).
\textbf{Right:} The \qty{32.1}{keV} \Kr S2 charge signal depends on drift time due to capture by electronegative impurities and can be described by an exponential function (blue). The S2 signals are corrected using this function towards the gate electrode position (orange).
}
\label{fig:correction_S2S1}
\end{figure}

Optical simulations show that the dependence of the S1~signal on the radial position is minimal and can be neglected~\cite{ABismarkThesis}. To compensate for the depth-dependent light collection efficiency caused by the geometry of the TPC, a linear fit is performed on the mean measured S1 area of \qty{32.1}{keV} decays of \Kr as a function of the corrected drift time, as seen in \cref{fig:correction_S2S1}, left. S1 signals are corrected with this function to achieve a homogeneous response, using the center of the drift region as reference position. The corrected signal is denoted~cS1.

Electronegative impurities such as oxygen or water can absorb electrons drifting towards the gate electrode. This reduces the observed S2 signal depending on the depth of an event. The exponential loss of electrons is described by the so-called electron lifetime. It is determined by performing Gaussian fits on the area distribution of the first S2 of \Kr events within drift time bins.
An exponential function is fit to the means of these Gaussians as a function of the drift time, as seen in \cref{fig:correction_S2S1}, right.
To compensate for the charge loss, the S2 signals are scaled towards the gate electrode using the electron lifetime to obtain the corrected S2 signal cS2.
Since this electron lifetime varies over time, this correction is performed for each run independently.
An electron lifetime of \qty{80}{\mus} was achieved during the measurements used for this work, which is about twice the maximal drift time.

\begin{figure}
\centering
\includegraphics{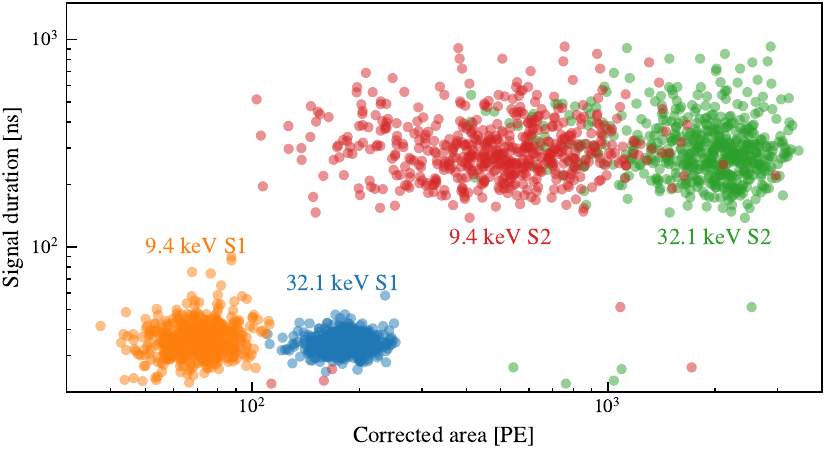}
\caption{
The four S1 and S2 signals of \Kr events inside the fiducial volume for an anode to gate electrode voltage difference of $\dV=\qty{4.4}{kV}$. The S2 signals (green and red) are wider and larger than the S1s (blue and orange).}
\label{fig:area_over_width_plot}
\end{figure}

After selecting and correcting the \Kr data, the four signals populate distinct regions in a space defined by the signal duration and the corrected signal area, as shown in \cref{fig:area_over_width_plot}. The duration is defined as the interval between the 25\% and 75\% quantiles of the peak.
The duration can be used to classify signals as S1 or S2. Further information on signal classification is given in~\cref{sec:classification}.

\subsection{Horizontal position reconstruction}
\label{sec:posrec}
The results of this work do not require radial fiducialization since \Kr events can be identified with minimal backgrounds and thanks to the small radial dependence of the TPC's S1 and S2 response.
For large rare-event detectors, however, horizontal ($xy$) position reconstruction is a critical feature to reduce backgrounds originating outside the TPC or in its walls.
In this section, we study the $xy$ position reconstruction power of the single-phase TPC, based on the distribution of the S2 light signal across the $1\times1$'' photomultipliers in the top array.

The reconstruction of the S2 signals' horizontal positions makes use of a deep feed forward neutral network, trained with the simulated light response. The neural network is built using TensorFlow~\cite{tensorflow}, accessed via the Keras API~\cite{keras}. The input layer of the neural net is given by the relative signals of the seven top array PMTs. This is connected to the output layer, consisting of the two horizontal positions $x$ and $y$, via two hidden layers with 64~nodes each.

\begin{figure}
\centering
\includegraphics{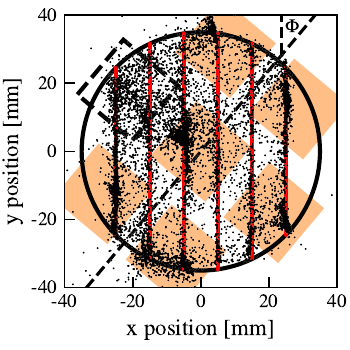}
\hfill
\includegraphics{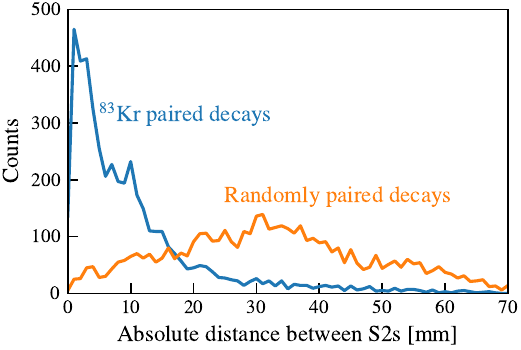}

\caption{
Horizontal position reconstruction in the single phase TPC. 
\textbf{Left:} Reconstructed event positions along the six anode wires (red lines) which were used to define the coordinate system. The square photocathodes of the top array PMTs are shown in the background (light orange). The line of symmetry of the PMTs (dashed black line) is rotated by $\Phi\approx\ang{40}$ with respect to the wires. The black circle denotes the reflective PTFE wall. The PMT located at $(x,y)=(-21,+18)$\,mm was not operational (dashed box).
\textbf{Right:} The distance between the reconstructed positions from each of the two decays of a \Kr nucleus has a median of \qty{7}{mm} (blue). This is compared to randomly paired events with a median distance of \qty{32}{mm} (orange).
}
\label{fig:posrec}
\end{figure}

The simulated light response for training is generated using GEANT4~\cite{GEANT4:2002zbu}. 20,000~events are randomly generated  along each of the six anode wires. Each event consists of 1000~individual photons, which are created in a cylinder of \qty{20}{\mum} radius around the \qty{10}{\mum} wire. This includes the full volume in which proportional scintillation in liquid xenon is expected~\cite{singlephase_tpc_kuger}. The relative orientation~$\Phi$ of the anode wires, which extend along the $y$-coordinate, to the top PMT array was not precisely measured and could only be reconstructed to between \ang{30} and \ang{50}. An angle $\Phi=\ang{40}$ is used to train the neural network, as indicated in~\cref{fig:posrec}, left. Studies showed negligible dependence of the neural network's performance on the angle used for training. One of the outer PMTs in the top array was not operational during data taking and is thus not used for the training.

The reconstructed positions of events acquired at all \dV are shown in \cref{fig:posrec}, left.
Although the \Kr decays are expected to be homogeneously distributed throughout the TPC, their reconstructed positions are clustered under the top PMTs. We attribute this to imperfections in the neural network, which could be because the simulated geometry and optics do not exactly match the real detector. An asymmetry is also visible in the distribution of reconstructed positions, which could be partly due to the non functional PMT. To quantify the reconstruction quality, the reconstructed positions of the S2s from the two \Kr decays are compared. Thanks to the high efficiency of tagging \Kr events and the short time between the two decays, these are known to come from the same position.
The distribution of the difference between the two reconstructed positions, where each decay is reconstructed separately, is shown in \cref{fig:posrec}, right. The median difference between the positions is \qty{7}{mm}, and is representative of the position reconstruction resolution. This is compared to the distance between randomly paired S2s, which are expected to be uncorrelated.
For these randomly paired decays, the median distance is \qty{32}{mm}, close to half the TPC diameter.
\section{Results}
\label{results}

Single-phase data of good quality could be acquired at eight anode-gate voltage differences~\dV between \qty{3.0}{kV} and \qty{4.4}{kV}, at two absolute gate and screening electrode voltages of \qty{-1}{kV} and \qty{-2}{kV}. At higher~\dV spurious light emission severely affected data taking. The drift field was kept constant for all measurements. The \qtylist{32.1; 9.4}{keV} decays of \Kr were used to measure the scintillation yield, energy resolution and other characteristics of the single-phase TPC.

\subsection{Electron drift velocity}

Using the drift time from the cathode~$T_\text{c}$ and gate electrode~$T_\text{g}$ and the cathode-gate distance of $\Delta z=\qty{69}{mm}$ at LXe temperature, the electron drift velocity can be determined as $v_\text{D}= \Delta z / (T_\text{c} - T_\text{g}) =\qty{1.75(3)}{mm/\us}$ at the drift field of \qty{473}{V/cm}. The uncertainty is dominated by the errors on $T_\text{c}$ and $T_\text{g}$.
This value agrees with other recent measurements~\cite{experiment_xebra,Jorg:2021hzu}.

\subsection{Secondary scintillation yield}
 
\begin{figure}
\centering
\includegraphics{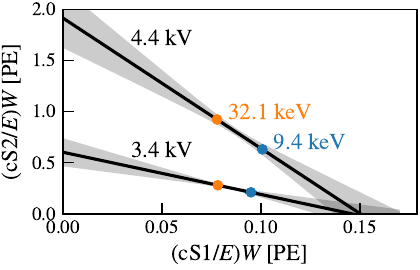}
\caption{Fit to the light and charge signals at $\dV=\qty{3.4}{kV}$ and \qty{4.4}{kV} to obtain the $g_1$ and $g_2$ gain factors. The shaded bands represent the uncertainty from the fit.
}
\label{fig:dokeplot}
\end{figure}

The equation 
\begin{align}
E = W \left(\frac{\mathrm{cS1}}{g_1} + \frac{\mathrm{cS2}}{g_2} \right)\label{eq:E_reconstr}
\end{align}
relates the energy~$E$ of the two krypton decays at \qty{32.1}{keV} and \qty{9.4}{keV} to the corrected S1 and S2 areas. The constant $W=\qty{13.7}{eV}$ is the average xenon excitation and ionisation energy~\cite{DahlThesis}. The gain factors $g_1$ and $g_2$, for the S1 and S2 signals, respectively, give the number of photoelectrons detected per quantum produced, after signal corrections are applied. They can be determined from a linear fit to the corrected signals cS1 and cS2 as shown in \cref{fig:dokeplot} for $\dV=\qty{3.4}{kV}$ and \qty{4.4}{kV}. 
Only \Kr events with decay times larger than \qty{600}{ns} are used to determine $g_1$ and $g_2$. For shorter decay times, the electrons from the first decay impact the local electrical field, resulting in a different charge yield for the second decay~\cite{Baudis:2013cca,result_short_kr_decays}.

\begin{figure}[b]
\centering
\includegraphics{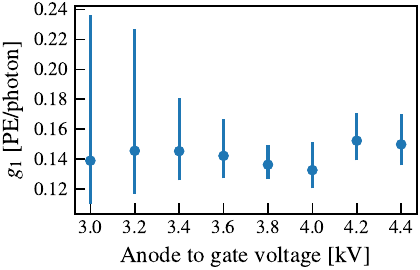}
\hfill
\includegraphics{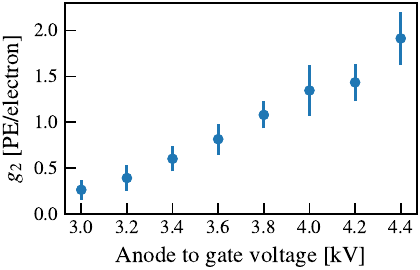}
\caption{The gain factors $g_1$ and $g_2$ for increasing anode-gate voltages~\dV.
Since the drift field is constant, $g_1$ (left) does not change with increasing \dV while $g_2$ (right) increases up to \qty{1.9(3)}{\pepel}.
}
\label{fig:g1g2_over_dV}
\end{figure}

The gain factors are determined independently for all studied voltages~\dV, as seen in \cref{fig:g1g2_over_dV}.
As expected, $g_1$ is independent of~\dV, as the light is produced in the active TPC target, unaffected by the anode voltage. The individual measurements are thus averaged to $g_1 =  \qty{0.142(8)}{\pepph}$.
This value is slightly higher than the \qty{0.122(2)}{\pepph} obtained in the dual-phase version of the TPC~\cite{experiment_xebra}. This is explained by the absence of the liquid-gas interface, which can cause total internal reflection and a second pass of the light through the gate electrode in dual-phase operation, and by the higher optical transparency of the single-phase anode.
The \dV-dependence of the charge gain $g_2$ is shown in \cref{fig:g1g2_over_dV}, right. It ranges from \qty{0.26(11)}{\pepel} at $\dV=\qty{3.0}{kV}$ to \qty{1.9(3)}{\pepel} at $\dV=\qty{4.4}{kV}$.

The analysis was repeated for \Kr events with drift times shorter than \qty{2.5}{\mus}, i.e., above the fiducial target selected above, to estimate the photon detection efficiency for S1 signals in a region close to the anode, where the proportional S2 signals are created.
Uncorrected signals were used: the S1 correction is not applied in order to measure the absolute local light yield and the S2 correction is not required as the high electron lifetime and the short drift time result in fewer than 4\% of electrons being absorbed by impurities.
Runs with \dV from \qtyrange{4.0}{4.4}{kV}  were grouped together to increase statistics.
Using the procedure above yields a photon-detection efficiency close to the anode region of $g_1^\text{a}=\qty{0.091(13)}{\pepph}$.
By dividing the S2 gain $g_2$ by this value, we can determine the detectable electroluminescence gain, or the number of detectable photons created per electron reaching the anode wire.
At the highest anode voltage $\dV=\qty{4.4}{kV}$, this is \qty{21(4)}{photons/electron}.

A model for the electroluminescence gain has been proposed by Aprile et al.~\cite{singlephase_tpc_elena}.
The charge gain $\Delta \nel$, for a step size~$\Delta r$ at a distance~$r$ from the center of an anode wire, relates the increase in the number of electrons~$\Delta\nel$ to the current number of electrons~$\nel$:
\begin{align}
\Delta \nel = \nel \Theta_0 \exp\left(-\frac{\Theta_1}{E(r; \dV, d_\text{w})- \Theta_2} \right) \Delta r. \label{eq:model_dne}
\end{align}
It is described by the three parameters $\Theta_{0,1,2}$, where $\Theta_2$ acts as a threshold field above which amplification is possible, and depends on the electric field $E(r; \dV, d_\text{w})$. The field is in turn a function of the anode-gate voltage and the wire diameter~$d_\text{w}$.
Using the values for $\Theta_i$ from~\cite{singlephase_tpc_elena}, the total charge multiplication factor is about~1.4 for the highest~\dV considered here.
The number of electrons is then used to calculate the number of photons $\Delta \nph$ generated in each step, where the parameters $\Theta_3$ and $\Theta_4$ describe the proportional scintillation per electron and $\Theta_4$ is the threshold for electroluminescence:
\begin{align}
\Delta \nph = \nel \Theta_3\left(E(r; \dV, d_\text{w}) - \Theta_4\right) \Delta r.\label{eq:model_dng}
\end{align}
Each formula is valid only when the electrical field is larger than the threshold, otherwise the gain in that step is zero. 

\begin{figure}
\centering
\includegraphics{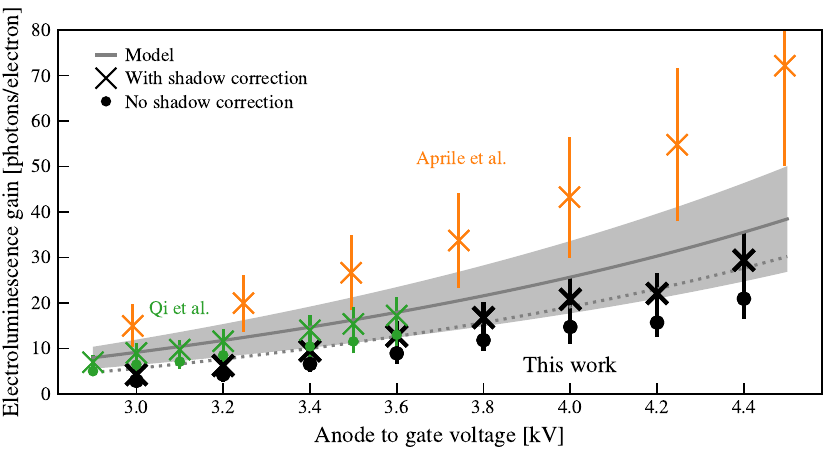}
\caption{
Comparison of our measured electroluminescence gain (black) to similar data from Qi et al.~\cite{radial_tpc_qi} (green) and Aprile et al.~\cite{singlephase_tpc_elena} (orange).
Crosses indicate the shadow corrected, emitted electroluminescence gain, dots indicate the uncorrected, detectable gain.
The data of Aprile et al. are based on a measurement using only the bottom PMT and are therefore considered to be shadow corrected.
The gain predicted by the model from~\cite{singlephase_tpc_elena}, with our electric field configuration, is also shown (solid gray line with uncertainty band).
Increasing the proportional scintillation threshold $\Theta_4$ in the model by 15\% yields a lower predicted gain (dotted gray line).
Error bars are shown for data both with and without the shadow correction.
}
\label{fig:comparision_elena}
\end{figure}

As the S2 light creation happens within a few micrometers to the anode wire~\cite{singlephase_tpc_kuger}, a fraction of the photons hits the wire and possibly escapes detection.
By including the geometrical coverage of the wire at each step in the model, the fraction of lost photons can be determined. This is done under the assumption of zero reflectivity and infinitely long wires.
Correcting for this, the detectable electroluminescence gain quoted above increases to an emitted electroluminescence gain of \qty{29(6)}{photons/electron} at $\dV=\qty{4.4}{kV}$.

A direct fit of the light production model given in \cref{eq:model_dne} and \cref{eq:model_dng} to our data was not possible because the data do not cover a sufficiently large \dV range to constrain the parameters. However, the $\Theta_i$ parameters for the \qty{10}{\mum} wire from~\cite{singlephase_tpc_elena} result in a model which is compatible with our data. The electric field $E(r;\dV,d_\text{w})$ needed for this model was obtained using finite element method simulations of the anode electrode installed in our TPC.

The electroluminescence yield from our single-phase TPC is shown in~\cref{fig:comparision_elena} and compared to the model.
Increasing $\Theta_4$ in \cref{eq:model_dng}, which represents the threshold for proportional scintillation in LXe, by 15\% to $\Theta_4=\qty{460}{kV/cm}$ yields a better match to our data. This value is close to the \qty{465}{kV/cm} found in a recent work on proportional scintillation in liquid xenon around microstrips~\cite{microstrips_wis}. 
For comparison, we also show the electroluminescence gains obtained by Qi et al.~\cite{radial_tpc_qi} and Aprile et al.~\cite{singlephase_tpc_elena}, also for \qty{10}{\um} wires. In the latter case, we use the conversion factor provided in the paper to determine the gain from the yield in photoelectrons.
Our shadow-corrected data are compatible with the original model as well as the data of Qi et al., albeit with a systematic offset. However, the measurements by Aprile et al.~show slightly higher gains. They also appear to be in conflict with their own model over this \dV range:
Their results cover a larger range of anode voltages and the measurements at higher \dV constrain the model and produce this tension.

\subsection{S2 resolution}

\begin{figure}
\centering
\includegraphics{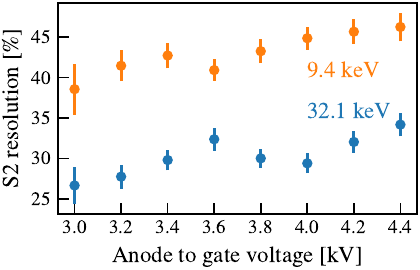}
\caption{
Single-phase S2 resolution for the different \dV and both \Kr decays. Although the signal size increases with increasing \dV, the resolution worsens.}
\label{fig:E_reconstr}
\end{figure}

The S2 resolution plays a role in the energy resolution of LXe TPCs and their ability to perform particle identification. Here, we study the impact of the single-phase technology on the S2 resolution. For both \Kr lines and all \dV values, the S2 area distributions are fitted by a Gaussian function with mean~$\mu$ and standard deviation~$\sigma$ to obtain the energy resolution $\sigma/\mu$ and its uncertainty
\begin{align}
\sigma^2_\frac{\sigma}{\mu} = \frac{\sigma^2_\sigma}{\mu^2} + \frac{\sigma^2 \sigma^2_\mu}{\mu^4} + \frac{-2\sigma \mathrm{cov}_{\mu\sigma}}{\mu^3}.
\end{align}
Here $\sigma_i$ denotes the uncertainty of the variable $i$ obtained from the fit.
The resolution as a function of \dV are shown in \cref{fig:E_reconstr}.
Over the \dV range studied here, the resolution of the \qty{32.1}{keV} decay's S2 deteriorates from $(27 \pm 2)$\% at $\dV=\qty{3.0}{kV}$ to $(34 \pm 2)$\% at \qty{4.4}{kV}.
The resolution of the \qty{9.4}{keV} decay is $(39\pm3)$\% at \qty{3.0}{kV}, reaching $(46 \pm 2)$\% at \qty{4.4}{kV}.
This trend could be caused by the increased electron multiplication and corresponding fluctuations.
According to the model from~\cite{singlephase_tpc_elena}, there is negligible electron multiplication at the lowest anode voltages, increasing to multiplication by a factor of about 1.4 at $\dV=\qty{4.4}{kV}$.
Previous results have also shown that the S2 resolution stops improving and eventually worsens with large enough \dV, although the threshold for this to happen varies slightly~\cite{singlephase_tpc_elena, microstrips_wis}.

\subsection{Duration of the secondary electroluminescence signals and electron diffusion}
\label{ssec:S2_width}

The duration of S2 signals directly impacts the ability of a TPC to separate interactions happening close together.
This can help to reduce backgrounds such as multiply-scattering neutrons in a dark-matter search.
Here we define the duration as the time between 25\% and 75\% of the total S2 area having been detected.

\begin{figure}
\centering
\includegraphics{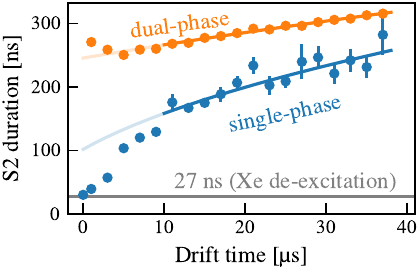}
\hfill
\includegraphics{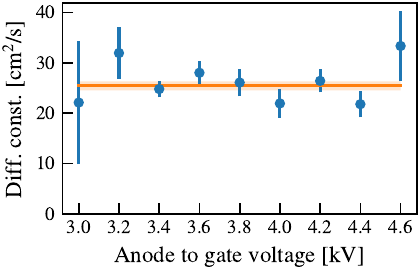}
\caption{
{\bf Left:} Comparison of the S2 durations measured in single-phase ($\dV=\qty{4.0}{kV}$) and dual-phase mode~\cite{experiment_xebra} as a function of the drift time.
A fit to the data is performed in the range from \qtyrange{10}{38}{\mus} to extract the electron diffusion constants.
The duration of the single phase S2 signals is dominated by diffusion for drift times above about \qty{5}{\us}, but approaches the xenon de-excitation timescale for shorter drift times. The deviation from the fit seen at short drift times is due to non-uniformity of the electric field near the gate electrode. {\bf Right:} The longitudinal diffusion constant $D_L$ does not depend on \dV, as expected.
}
\label{fig:width_fit}
\end{figure}

The duration of the single-phase~S2 signal is dominated by diffusion and, for short drift times, the \qty{27}{ns} triplet de-excitation time of the Xe$_2^*$ excimers.
An example for $\dV=\qty{4.0}{kV}$ is shown in \cref{fig:width_fit}, left, together with the data from operating the same TPC in dual-phase mode~\cite{experiment_xebra}. The single-phase S2 signals are significantly shorter than the dual-phase signals, for which the multi-millimeter path over which the secondary light is created contributes significantly to their duration.
For low drift times the duration of single-phase S2s approaches the xenon de-excitation timescale. Only at drift times beyond the maximal drift time in this TPC would the duration be dominated by electron diffusion in both types of detector and then be comparable.

To obtain the longitudinal electron diffusion constant~$D_L$, the measured S2 duration~$w$ is plotted against the event's drift time and fitted by the diffusion formula
\begin{align}
w = \sqrt{\frac{2D_Lt}{v^2_\mathrm{drift}} \ + \ w^2_0}.
\end{align}
The fit is restricted to the central part of the TPC, with drift times from \qtyrange{10}{38}{\mus}. This is the region with a homogeneous drift field and therefore a constant drift velocity. For the single-phase data, this fit was performed for all \dV; the results are shown in \cref{fig:width_fit}, right. As expected, the diffusion constant does not depend on \dV.
The average longitudinal diffusion constant $D_L=\qty{25.4(9)}{cm^2/s}$ at the field of \qty{473}{V/cm} is compatible with the \qty{25.7(45)}{cm^2/s} measured by Njoya et~al.~at \qty{500}{V/cm}~\cite{diffusion_njoya}, but higher than the \qty{19.5(6)}{cm^2/s} obtained by Hogenbirk et~al.~at \qty{490}{V/cm}~\cite{result_diffusion_hogenbirk}.

\subsection{S1 and S2 signal identification}
\label{sec:classification}
\begin{figure}[ht]
\centering
\includegraphics{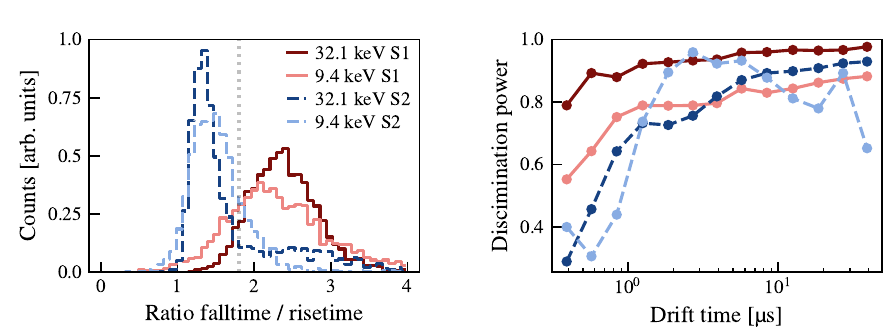}
\caption{\textbf{Left:} The ratio $\cal R$ of fall-time to rise-time for S1s and S2s at an anode voltage $\dV=\qty{4.4}{kV}$ for \Kr decays with a drift time between \qty{1}{\us} and \qty{2}{\us}. A ratio of 1.8 was used to split the regimes (dotted line).
\textbf{Right:} The discrimination power increases with the drift time and, in the case of S1s, the signal size. Shown here is data for the rather low $\dV=\qty{3.4}{kV}$; the discrimination power does not seem to depend on \dV.
}
\label{fig:risefalltime}
\end{figure}

For the results in this work, the S1 and S2 signals can be distinguished based on their time order within the unique \Kr signature.
For more general applications, however, a different classification method is required, that does not depend on the time ordering. This can be achieved by exploiting the different underlying production processes, leading to different S1 and S2 signal shapes.
S1s show a very steep rise followed by an exponential fall, while S2s have a more symmetric shape, where rise and fall times are similar.
Here, we use the fall-time to rise-time ratio
\begin{equation}
    {\cal R}=\frac{t_{90}-t_{50}}{t_{50}-t_{10}},
\end{equation}
where $t_p$ is the time by which $p$~percent of the peak's area has been recorded.
\cref{fig:risefalltime}, left shows the distribution of the ratio $\cal R$ for the four \Kr decays, and the value ${\cal R} = 1.8$ used to distinguish S1s from S2s. The discrimination power between S1s and S2s, defined by the fraction of peaks assigned the correct category, is shown in \cref{fig:risefalltime}, right.
It rises from around 40\% at short drift times to above 80\% for drift times longer than about \qty{5}{\us}, where diffusion plays a greater role. It also improves with increasing signal size. This figure shows data for the rather low $\dV=\qty{3.4}{kV}$, but the discrimination is not significantly affected by \dV.

\section{Conclusions}
\label{discussion}

Using proportional scintillation in LXe to measure the charge signal in TPCs could help solving some of the most pressing construction and operation challenges, mainly related to the electrodes and the TPC's long-term stability. It would mitigate delayed electron extraction, allow for new analysis methods such as electron counting~\cite{singlephase_tpc_kuger}, and enable different detector designs such as the radial TPC~\cite{Ye:2014gga,radial_tpc_qi}. 

In this work, we have demonstrated that a small-scale, single-phase TPC can be successfully operated, characterized and analyzed analogously to dual-phase TPCs.
The design of our TPC closely follows the dual-phase design employed successfully in a large number of dark matter experiments: a cylindrical TPC with approximately 1:1 aspect ratio and light readout above and below the target. Since the TPC used for this study was previously operated in dual-phase mode~\cite{experiment_xebra}, a direct comparison between the two signal generation processes was possible. The high electric fields required to generate proportional scintillation in the liquid phase were established around thin gold-plated tungsten anode wires of \qty{10}{\mum}~diameter, by establishing a voltage difference \dV between the anode and the gate electrodes. Proportional scintillation from \Kr calibration events was observed for $\dV\ge\qty{3.0}{kV}$. Above \qty{4.4}{kV} spontaneous light emission in the TPC prevented stable operation, as also observed in~\cite{radial_tpc_wei}. At the highest stable \dV of \qty{4.4}{kV}, a scintillation gain of $g_2=\qty{1.9(0.3)}{\pepel}$ was achieved. This corresponds to an electroluminescence gain of \qty{29(6)}{photons/electron}, after correcting for shadowing by the anode wire. The observed \dV-dependence is comparable to that reported in~\cite{radial_tpc_qi}. It is also compatible with the model proposed in~\cite{singlephase_tpc_elena}, however, it might hint at a slightly higher electroluminescence threshold.

The maximum proportional scintillation gain factor $g_2$ achieved in the single-phase TPC is lower than the \qty{5.49(5)}{\pepel} achieved during dual-phase operation, with an extraction field (in the LXe) of \qty{2.8}{kV/cm}~\cite{experiment_xebra}. It is also significantly lower than the gains achieved in large dual-phase TPCs searching for dark matter, with $g_2$ values ranging from around \qty{17}{\pepel} in XENONnT~\cite{result_xenonnt_lowER} to almost \qty{60}{\pepel} in LZ~\cite{LZ:2023poo}. Since the S2 size impacts the rejection of electronic recoil backgrounds, higher single-phase $g_2$ gains would be needed to consider this technology as an alternative to the dual-phase TPC.
Achieving higher gains will require improving our understanding of the spontaneous light emission and ways in which it can be reduced.
Using larger diameter anode wires could allow larger gains to be reached for a given surface field, possibly reducing this light emission.
For single-phase technology to be used by a large-scale detector, additional studies will be needed to show that thin wires can stably be installed and operated with lengths of several meters.

Using data from our single-phase TPC, we measured the electron drift velocity and the longitudinal electron diffusion constant at a drift field of \qty{473}{V/cm}, with values in agreement with measurements in dual-phase TPCs. We have shown that single phase S2~signals can be used for three-dimensional position reconstruction and target fiducialization. Their duration is dominated by electron diffusion. This leads to narrower peaks at shorter drift times, potentially enabling new analysis techniques such as electron counting, and improving the identification of multiple S2 peaks in an event, which directly benefits background rejection in rare event searches.

\acknowledgments

This work was supported by the European Research Council (ERC) grant No.~724320 (ULTIMATE). We thank the teams of the mechanical and electronics workshops of the Institute of Physics, Freiburg, for their continuous support. Finally, we thank all the Bachelor students and interns who contributed to commissioning and operation of the detector, as well as data analysis.

%\bibliography{xebra_paper}

\providecommand{\href}[2]{#2}\begingroup\raggedright\endgroup

\end{document}